\documentclass[11pt]{article}
\usepackage{amsfonts,amscd,amsmath, amssymb,graphicx}
\def\ep{\text{e}}

\def\zt{z_{\text{\tiny T}}}
\def\T{T_c}

\def\zt{z_{\text{\tiny T}}}

\def\cs{C_{\text{\tiny S}}}

\title{ Some Thermodynamic Aspects of Pure Glue, Fuzzy Bags and Gauge/String Duality}
\author{Oleg Andreev \thanks{Also at Landau Institute for Theoretical Physics, Moscow, Russia. Email address: andre@itp.ac.ru}\\\\
{\it Max-Planck Institut f\" ur Physik, F\" ohringer Ring 6,} \\
{\it 80805 M\" unchen, Germany;}
}
\date{}

\begin{document}

\vspace{-8cm}
\maketitle
\begin{abstract}
The thermodynamic properties of a $SU(3)$ gauge theory without quarks are calculated using a string formulation for 
$1.2\,\T\leq T\leq 3\,\T$. The results are in good agreement with the lattice data. We also comment on $SU(N)$ gauge theories.
 \end{abstract}

\vspace{-12.5 cm}
\begin{flushright}
MPP-2007-72\\
\end{flushright}

\vspace{12 cm}

%___________________________________________________________________________________________________

\section{Introduction}
High energy nucleus-nucleus collisions provide the means of creating nuclear matter in conditions of extreme temperature and density. In 
particular, the system undergoes a transition from a state of nucleons containing bound quarks and gluons to a state of deconfined quarks 
and gluons. This state was originally given the name Quark Gluon Plasma. However, the results at RHIC indicate that instead of behaving like a 
weakly coupled gas of free quarks and gluons, the matter created in heavy ion collisions behaves like a strongly coupled liquid.\footnote{For 
recent reviews, see \cite{rhic}.} Thus, there is a need for new approaches to strongly coupled gauge theories.

Until recently, the lattice formulation was a unique theoretical tool to deal with strongly coupled gauge theories. The subject has taken an 
interesting turn with Maldacena duality \cite{malda0}. One of the implications is that it resumed interest in finding a string 
description of strong interactions. Although the original proposal was for conformal theories, various modifications have been found that 
produce gauge/string duals with a mass gap, confinement, and supersymmetry breaking \cite{malda}.

In this paper we address some issues of thermodynamics of $SU(3)$ pure gauge theory in a dual formulation. Clearly, finding the dual from 
first principles of string theory is beyond of our ability. Instead, we attempt the inverse problem and use our knowledge of some 
phenomenologically successful five-dimensional models of AdS/QCD. 

Before proceeding to the detailed analysis, let us set the basic framework. We consider the following ansatz for the 10-dimensional background 
geometry which turns out to be applicable for the temperature range $1.2\,\T\leq T\leq 3\,\T$ \footnote{The lower limit is 
chosen to keep the system out of the critical regime. As we will discuss below, the upper limit is determined by consistency rather 
than perturbation theory.}

\begin{equation}\label{metricT}
ds^2
=\frac{R^2}{z^2} H
\left(fdt^2+d\vec x^2+f^{-1}dz^2\right)+H^{-1}g_{ab}d\omega^ad\omega^b
\,,
\qquad
f=1-\Bigl(\frac{z}{\zt}\Bigr)^4
\,,\quad 
H=\ep^{\frac{1}{2}cz^2}
\,.
\end{equation}
Here $\zt=1/\pi T$. It is a deformed product of the Euclidean $\text{AdS}_5$ black hole and a 5-dimensional sphere (compact space) 
whose coordinates are $\omega^a$. The deformation is due to a $z$-dependent factor $H$. Such a deformation is crucial for breaking 
conformal invariance of the original supergravity solution and introducing $\Lambda_{\text{\tiny QCD}}$. 

Apart from the language of 10-dimensional string theory, there is a more phenomenological way to attack QCD. This approach 
called AdS/QCD deals with a five-dimensional effective description and tries to fit it to QCD as much as possible. For our model, 
its AdS/QCD cousin can be obtained by discarding the compact space in \eqref{metricT}. 

At $T=0$, then what we get is the slightly deformed $\text{AdS}_5$ metric. Such a deformation is notable. The point is that in this background 
linearized Yang-Mills equations are effectively reduced to a Laguerre differential equation. As a result, the spectrum turns out to be like that 
of the linear Regge models \cite{regge,oa}. This fact allows one to fix the value of $c$ from the $\rho$ meson trajectory. It is of order \cite{oa}

\begin{equation}\label{c}
c\approx 0.9\,\text{GeV}^2
\,.
\end{equation}
We will assume that the value of $c$ is universal and is therefore valid for the world without quarks too. In addition, this AdS/QCD model 
provides the phenomenologically acceptable heavy quark potentials as well as  the value of the gluon condensate \cite{az1,az4}.

At finite $T$, the model provides the spatial string tension of pure gauge theory \cite{az2}. The agreement with the lattice data is very good for 
temperatures lower than $2.5$-$3\,\T $. Due to this reason we set the upper bound on $T$ in \eqref{metricT}. Moreover, the model describes 
in a qualitative way a heavy quark-antiquark pair and the expectation value of the Polyakov loop \cite{az3}. 

Thus, there are reasons to believe that the model \eqref{metricT} is a good approximation  for a string dual to a pure gauge theory.

\section{Thermodynamics}
%___________________________________________________________________________________________________
\subsection{The Entropy Density}
One of the bedrocks of gauge/string (gravity) duality is a conjecture that the entropy of gauge theories is equal to the Bekenstein-Hawking 
entropy of their string (gravity) duals \cite{malda}. As known, the Bekenstein-Hawking entropy is proportional to an (8-dimensional) 
area of the horizon. We can now ask whether the five-dimensional framework (AdS/QCD) is an adequate approximation at this point. In general, 
the answer is no. There is a contribution from the compact space that might be relevant. 

The metric \eqref{metricT} has the horizon at $z=\zt$. Therefore, the temperature dependence of the entropy density is\footnote{We take a 
constant dilaton.}

\begin{equation}\label{entropy}
s(T)=s_0\, T^3\exp\Big\{-\frac{1}{2}\frac{\T^2}{ T^2}\Big\}
\,,
\end{equation}
where $s_0$ is a factor independent of temperature. In this formula $\T$ is given by\footnote{In the following section we will see that $\T$ can be 
thought as a critical temperature.}

\begin{equation}\label{Tc}
\T=\frac{1}{\pi}\sqrt{c}
\,.
\end{equation}

It follows from \eqref{entropy}  that the entropy density can be represented as a series in powers of $\frac{1}{T^2}$ 
with the leading $T^3$ term

\begin{equation}\label{entropy-series}
s(T)=s_0\,T^3\sum_{n=0}^{\infty}a_n\tau^n
\,,\qquad
\tau=\frac{\T^2}{T^2}
\,,
\end{equation}
where $a_n=\frac{(-)^n}{2^n n!}$.

For future use, we define the truncated model by keeping the two leading terms in \eqref{entropy-series}. We have

\begin{equation}\label{entropy-truncated}
s_{\text{\tiny tr}}(T)=s_0\,T^3\Bigl(1-\frac{1}{2}\tau\Bigr)
\,.
\end{equation}

%____________________________________________________________________________________________________
\subsection{The Pressure}
\subsubsection{Fuzzy Bags}

Recently, it has been suggested by Pisarski that for the temperature range $T_{\text{\tiny max}}<T<T_{\text{\tiny pert}}$ the pressure in 
QCD with quarks is given by a series in powers of $\frac{1}{T^2}$ times the ideal $T^4$ term \cite{pisarski}. Explicitly,

\begin{equation}\label{qcd-pressure}
p_{\text{\tiny QCD}}(T)\approx  f_{\text{\tiny{pert}}}T^4-B_{\text{\tiny fuzzy}}T^2-B_{\text{\tiny MIT}}+\dots
\,.
\end{equation}
It was called a fuzzy bag model for the pressure. So, $B_{\text{\tiny MIT}}$ stands for the MIT bag constant. $T_{\text{\tiny max}}$ is 
close to a critical temperature $\T$ (or some approximate $''\T''$ for a crossover). A small difference between $\T$ and 
$T_{\text{\tiny max}}$ may vary with the model. $T_{\text{\tiny pert}}$ is set by perturbation theory such that it is applicable only 
for temperatures higher than $T_{\text{\tiny pert}}$. 

For pure glue, Pisarski argued, based on lattice simulations of \cite{karsch}, that \eqref{qcd-pressure} reduces to 

\begin{equation}\label{pis-pressure}
p(T)\approx f_{\text{\tiny{pert}}}\left(T^4-\T^2 T^2\right)
\,.
\end{equation}
This means that $B_{\text{\tiny fuzzy}}= f_{\text{\tiny{pert}}}\T^2$ and $B_{\text{\tiny MIT}}$ is much smaller than 
the first two terms.  So, the pressure is a sum of just two pieces. Note that an important consequence of \eqref{pis-pressure} is that 
the pressure (nearly) vanishes at $T=\T$.  

%_______________________________________________________________________________________________
\subsubsection{String Dual}

Given the entropy density as a function of $T$, in the homogeneous case one can find the temperature dependence of the pressure 
by integrating $\frac{d p}{d T}=s$.\footnote{In what follows, we  consider the homogeneous case.} From \eqref{entropy-series}, we get 

\begin{equation}\label{pressure}
p(T)=\frac{1}{4}s_0 T^4
\Bigl(1-\tau-\frac{1}{4}\tau^2\ln\tau-b\,\tau^2+
\sum_{n=3}^{\infty}b_n\tau^n\Bigr)
\,,
\end{equation}
where $b$ is an integration constant and $b_n=\frac{2 a_n}{2-n}$.

The final topic to be considered here is whether the proposal of Pisarski is reasonable in the model under consideration. The two leading terms in 
\eqref{pressure} look similar to those of \eqref{pis-pressure}. So, we find that the critical temperature is given by $\T$. A simple estimate then 
gives\footnote{We use \eqref{c} for this estimate.}

\begin{equation}\label{Tc-est}
T_c\approx 300\,\text{MeV}
\,.
\end{equation}
In SU(3) pure gauge theory the critical temperature is of order $270$ MeV. So, the agreement is not bad at this point.

Let us now use $p(\T)=0$ to determine the integration constant. As a result, we have

\begin{equation}\label{b}
b=\sum_{n=3}^{\infty}b_n\approx 0.039
\,.
\end{equation}
The value of $b$ is indeed small compared to the coefficients in front of the two leading terms. Thus, the agreement is very satisfactory at 
this point.

To complete the picture, we present the results of numerical calculations. We split the series \eqref{pressure} into two pieces, the first 
containing the two leading terms, and the second presenting the rest. Then we define\footnote{Note that the 
truncated model \eqref{entropy-series} can be derived from $p_1$ times $\frac{1}{4}s_0\,T^4$.}

\begin{equation}
p_1(T)=1-\tau
\,,
\qquad
p_2(T)=-\frac{1}{4}\tau^2\ln\tau-b\,\tau^2+
\sum_{n=3}^{\infty}b_n\tau^n
\,.
\end{equation}
For the sake of simplicity, we have omitted the overall factor $\frac{1}{4}s_0\,T^4$. The values of $p_1$ and $p_2$  can be read off of Fig.1.
%__________________________________________________________________________________________
\begin{figure}[htbp]
\begin{center}
\includegraphics[width=5cm]{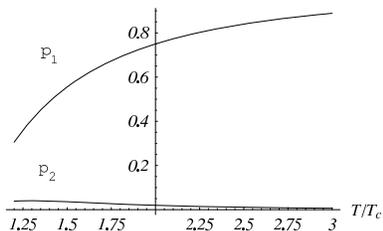}
\caption{\small{Values of $p_1$ and $p_2$ versus the ratio $\frac{T}{T_c}$.}}
%\label{fig:graph1}
\end{center}
\end{figure}
%______________________________________________________________
We see that at $T\approx 1.2\,\T$ the value of $p_2$ is one order of magnitude smaller than that of $p_1$. 
Above $1.2\,\T$ the value of $p_1$ increases, while $p_2$ decreases and becomes negligible for $T\gtrsim 2\,\T$. Thus, 
$p_1(T)$ provides a reliable approximation whose error is less then 10\% for the pressure. 

In sum, the truncated model which is equivalent to the proposal of Pisarski is valid with accuracy better than $10\%$.

%______________________________________________________________________________________________________
\subsection{The Speed of Sound}
Having derived the entropy density, we can easily obtain the speed of sound. For the model of interest, we have

\begin{equation}\label{cs}
\cs^2(T)=\frac{s}{Ts'}
=\frac{1}{3}\Bigl(1+\frac{1}{3}\tau\Bigr)^{-1}
\,.
\end{equation}
For completeness, we also present the result obtained for the truncated model \eqref{entropy-truncated}. In this case 
\eqref{cs} is replaced by 

\begin{equation}\label{cs2}
\cs^2(T)=\frac{1}{3}\Bigl(
1-\frac{1}{2}\tau\Bigr)\Bigl(1-\frac{1}{6}\tau\Bigr)^{-1}
\,.
\end{equation}
Note that $\cs$ is independent of $s_0$. Thus, we do not have any free fitting parameter at this point. 

We close the discussion of the speed of sound by comparing the results with those of lattice simulations.\footnote{The recent data 
of \cite{gavai} have large error bars. So, it is impossible to say how precisely the results fit. } The curves are 
shown in Fig.2.

%________________________  f - 2  __________________________________
%
%\vspace{.6cm}
\begin{figure}[htbp]
\begin{center}
\includegraphics[width=5.75cm]{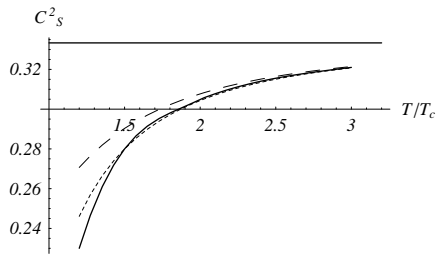}
\caption{\small{The square of the speed of sound versus $\frac{T}{T_c}$. The upper and lower dashed curves correspond to \eqref{cs} and  
\eqref{cs2}, respectively. The solid curve represents the result of the extrapolation to the continuum limit for lattice simulations \cite{karsch}. 
The solid horizontal line is the usual AdS/CFT result with the value $\frac{1}{3}$.}}
%\label{fig:graph1}
\end{center}
\end{figure}
%______________________________________________________________
%\vspace{-0.3cm}
From the Figure we can see that the model of interest is in very good agreement with the lattice for $T\gtrsim 1.7\,\T$, while 
near $1.2\,\T$ the discrepancy is of order $15$ percent. The agreement between the truncated model and the lattice is spectacular. The 
maximum discrepancy occurred at $T=1.2\,\T$ is of order $6$ percent.

\subsection{The Gluon Condensate at Finite Temperature}
We will next describe the gluon condensate at finite temperature.\footnote{Although the literature on the gluon condensate is very vast, 
to our knowledge, there are no reliable results for the temperature range $1.2\,\T\leq T\leq 3\,\T$ except those coming from lattice simulations.} It is 
obtained from the trace anomaly of the energy-momentum tensor \cite{anomaly}. We have

\begin{equation}\label{G2T-L}
G_2(T)=G_2+4p-Ts
\,,
\end{equation}
where $G_2$ is the condensate at zero temperature. 

Unlike the speed of sound, the condensate depends on the parameter $s_0$. There are two different ways to fix its value which fortunately 
yield very similar results. The first is to fit the interaction measure $(\epsilon-3p)/T^4$ as it follows from 
\eqref{entropy-series} and \eqref{pressure} to the lattice data of \cite{karsch} at some normalization point $T_n$. As result, we get 

\begin{equation}\label{s0}
s_0=6.8\pm 0.3
\,.
\end{equation}
At first glance it may seem curious that  the result is almost independent of the normalization point. As we will see in a moment, this is indeed 
the case.

The second is to match the coefficient in front of the $T^4$ term in \eqref{pressure} with that of the bag model \cite{mit}. For $SU(N)$ (pure) 
gauge theory, the latter is simply $\frac{N^2-1}{45}\pi^2$. At $N=3$, we find 

\begin{equation}\label{s0Bag}
s_0=\frac{32}{45}\,\pi^2\approx 7.0
\,
\end{equation}
that is really the same as \eqref{s0}.

Having determined the value of $s_0$, we can now write down the expression for the condensate. Combining \eqref{G2T-L}, and 
\eqref{entropy-series} and \eqref{pressure}, we get

\begin{equation}\label{G2T}
G_2(T)=-s_0\,T^4\biggl(\frac{1}{2}\tau+\frac{1}{4}\tau^2\ln\tau+ g\tau^2+\sum_{n=3}^{\infty}(a_n-b_n)\tau^n\biggr)
\,,
\end{equation}
where $g=\frac{1}{8}+b-\frac{k}{s_0}$. Note that the condensate at zero temperature $G_2=k\T^4$ has been included in the $\tau^2$ term. 
For the background geometry \eqref{metricT}, the estimate of \cite{az4} gives $k\approx 1.20$. Interestingly, the value of $g$ turns out to 
be small. For $s_0=6.8$ it is of order $-0.01$.

In Fig.3 we have plotted the gluon condensate in units of $\T^4$ as a function of the ratio $\frac{T}{\T}$. 

%________________________  f - 3  __________________________________
%
%\vspace{.6cm}
\begin{figure}[htbp]
\begin{center}
\includegraphics[width=5.75cm]{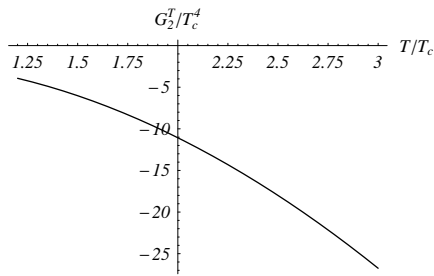}
\caption{\small{The gluon condensate in units of $\T^4$ versus $\frac{T}{\T}$. Here $s_0=6.8$.}}
%\label{fig:graph1}
\end{center}
\end{figure}
%______________________________________________________________
%\vspace{-0.3cm}

We conclude the discussion with a couple of comments:
\newline (i)  The expression for the gluon condensate is cumbersome and difficult of any practical use. We should therefore seek a simpler
(nearly equal) expression. What we already know is that the truncated model is a good approximation for the pressure and the speed of sound. So, 
it is reasonable to use this option. We can check it by the same arguments that we used in the case of the pressure. To this end, we split the 
series \eqref{G2T} into two pieces and define 

\begin{equation}
g_1(T)=-\frac{1}{2}\tau
\,,
\qquad
g_2(T)=-\frac{1}{4}\tau^2\ln\tau-g\tau^2+
\sum_{n=3}^{\infty}(b_n-a_n)\tau^n
\,.
\end{equation}
For simplicity, we have omitted the overall factor $s_0T^4$.

The values of $g_1$ and $g_2$ can be read off of Fig.4. 
%________________________  f - 4  __________________________________
%
%\vspace{.6cm}
\begin{figure}[htbp]
\begin{center}
\includegraphics[width=5.25cm]{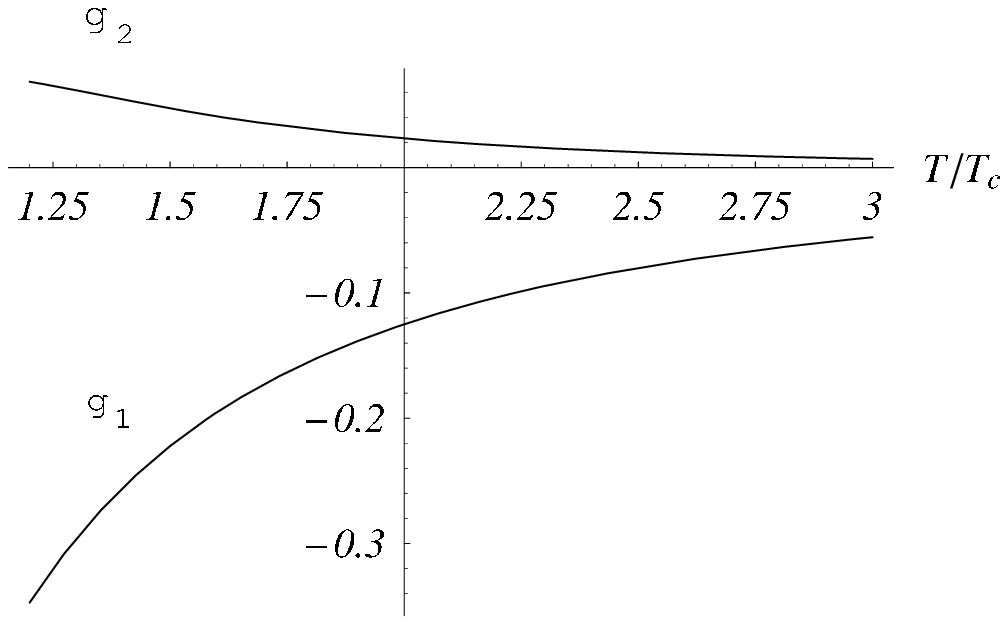}
\caption{\small{Values of $g_1$ and $g_2$ versus the ratio $\frac{T}{\T}$.}}
%\label{fig:graph1}
\end{center}
\end{figure}
%______________________________________________________________
%\vspace{-0.3cm}
We see that the value of $g_2$ is approximately $15\%$ of $g_1$. Thus, in the temperature range under consideration 
we may approximate the infinite series \eqref{G2T} by $g_1$. Finally, the gluon condensate takes the form predicted by the truncated model

\begin{equation}\label{G2T-app}
G_2(T)\approx -\frac{s_0}{2}\,\T^2 T^2
\,.
\end{equation}

\noindent (ii) Using \eqref{entropy-series} and \eqref{pressure}, one can easily find the expression for the interaction measure. It is 

\begin{equation}\label{measure}
\frac{\epsilon -3p}{T^4}=
s_0\,\biggl(\frac{1}{2}\tau+\frac{1}{4}\tau^2\ln\tau+
\Bigl(b+\frac{1}{8}\Bigr)\tau^2+\sum_{n=3}^{\infty}(a_n-b_n)\tau^n\biggr)
\,.
\end{equation}
The truncated model provides a simpler expression of the measure

\begin{equation}\label{measure-trunc}
\frac{\epsilon -3p}{T^4}=\frac{s_0}{2}\,\tau
\,,
\end{equation}
as expected. 

In Fig.5 we have plotted the interaction measure as a function of the ratio $\frac{T}{\T}$. As can be seen from the Figure, the agreement with 
the lattice data is very satisfactory. An important observation is that varying $s_0$ over the range \eqref{s0} has a little effect.

%________________________  f - 5  __________________________________
%
%\vspace{.6cm}
\begin{figure}[htbp]
\begin{center}
\includegraphics[width=5.75cm]{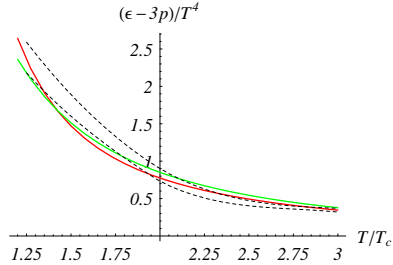}
\caption{\small{The interaction measure $(\epsilon-3p)/T^4$ versus $\frac{T}{\T}$. The red and green curves correspond to 
\eqref{measure} and \eqref{measure-trunc}, respectively. The lattice data of \cite{karsch} lie between the two dashed lines. Here $s_0=6.8$.}}
%\label{fig:graph1}
\end{center}
\end{figure}
%______________________________________________________________
%\vspace{-0.3cm}

\section{Concluding Comments}

(i) The model we have proposed predicts the entropy density as a series in $\frac{1}{T^2}$. It differs from the proposal of Pisarski  \cite{pisarski} 
by having a term $\ln T$ in the pressure. However, in the pure glue case the $\ln T$ term turns out to be subdominant. 

\noindent (ii) Interestingly enough, the spatial string tension calculated within the AdS/QCD cousin of \eqref{metricT} is given by \cite{az2}

\begin{equation}\label{sigma-s}
\sigma_s=\sigma\,\frac{\ep^{\tau-1}}{\tau}
\,,
\end{equation}
where $\sigma$ is the string tension at zero temperature. Then, from \eqref{sigma-s}, we see that the spatial tension can be written as a
series in powers of $\frac{1}{T^2}$ times $T^2$. Note that unlike the cases considered in section 2 the first two terms of the series do not 
provide a reasonable approximation. 

\noindent (iii) Can one think of the model \eqref{metricT} as a string dual to a $SU(N)$ pure gauge theory? We will be exploring the 
consequences of assuming that the pressure vanishes at $T=\T$.\footnote{The parameter $c$ is 
now dependent of $N$, so \eqref{c} is not applicable.}

This assumption leads to the same expression for the pressure as \eqref{pressure} with $b$ defined by \eqref{b}. The overall constant $s_0$ 
is fixed from the $T^4$ term. Fitting the bag model, we have 

\begin{equation}\label{s0BagN}
s_0=\frac{4\pi^2}{45}\left(N^2-1\right)
\,.
\end{equation}

Clearly, the analysis of section 2 is not sensitive to $N$. So, the conclusion we draw is that the truncated model which is equivalent to 
\eqref{pis-pressure} is valid with accuracy better than $10\%$. 

Moreover, we can obtain a formula for the pressure normalized by the leading term $p_0=\frac{1}{4}s_0T^4$. It is 

\begin{equation}\label{pressure-nor}
\frac{p}{p_0}(\tau)=
\Bigl(1-\tau-\frac{1}{4}\tau^2\ln\tau-b\,\tau^2+
\sum_{n=3}^{\infty}b_n\tau^n\Bigr)
\,.
\end{equation}
Thus our model predicts that the ratio is a function of $\frac{T}{\T}$.\footnote{Strictly speaking, it is a function of $\frac{\T^2}{T^2}$. }
It does not explicitly depend on $N$. At this point it is worth noting that in addition to $N=3$ the prediction is also supported by lattice simulations 
for $N=4$ and $N=8$ \cite{teper}.

\noindent (iv) We can gain some understanding of the $N$ dependence of a parameter $\mathfrak{ g}=\frac{R^2}{\alpha'}$. Here $\alpha'$ is the 
usual string parameter coming from the Nambu-Goto action. 

The lattice data are well fitted by \cite{teper2}
\begin{equation}\label{teper}
\frac{\T}{\sqrt{\sigma}}=0.596+\frac{0.453}{N^2}
\,,
\end{equation}
where $\sigma$ is the string tension at zero temperature. For the AdS/QCD cousin of \eqref{metricT} it is given by \cite{az1}  

\begin{equation}\label{tension}
\sigma=\mathfrak{g}\frac{e}{4\pi}c
\,.
\end{equation}

Combining \eqref{teper}, and \eqref{Tc} and \eqref{tension}, we learn

\begin{equation}\label{g}
\mathfrak{g}=\frac{4}{\pi e}\left(0.596+\frac{0.453}{N^2}\right)^{-2}
\,.
\end{equation}
A simple algebra shows that $\mathfrak{g}$ is a slowly varying function of $N$. It takes values between $0.93$ at $N=2$ and $1.32$ at 
$N=\infty$. 

For $N=3$, $\mathfrak{g}$ is approximately equal to $1.12$. It is interesting to compare this value with the estimate of \cite{az1}. 
The latter was made by using the Cornell potential. The result is $\mathfrak{g}\approx 0.94$. The estimates are relatively close. This
might be a hint that $\mathfrak{g}$ is also a slowly varying function of a number of quarks. 

%___________________________________________________________________________________________
\vspace{.25cm}

{\bf Acknowledgments}

\vspace{.25cm}
This work was supported in part by DFG and Russian Basic Research Foundation Grant 05-02-16486. 
We are grateful to R.D. Pisarski for correspondence and to V.I. Zakharov and P. Weisz for many stimulating discussions.

%__________________                      R E F S                    ______________________

%____________________________________________________________________

\small
\begin{thebibliography}{99}
\bibitem{rhic}
M.J. Tannenbaum, Heavy Ion Physics at RHIC, nucl-ex/0702028;\\
 E. Shuryak, Strongly Coupled Quark-Gluon Plasma: The Status Report; hep-ph/0703208.
\bibitem{malda0}
J.M. Maldacena, Adv.Theor.Math. Phys. {\bf 2}, 231 (1998).
\bibitem{malda}
O. Aharony, S.S. Gubser, J.M. Maldacena, H. Ooguri, and Y. Oz, Phys. Rept. {\bf 323}, 183 (2000).
\bibitem{regge}
R.R. Metsaev, IIB supergravity and various aspects of light cone formalism in AdS space-time, hep-th/0002008;\\
A. Karch, E. Katz, D.T. Son, and M.A. Stephanov, Phys.Rev.D {\bf74}, 015005 (2006).
\bibitem{oa}
O. Andreev, Phys.Rev. D {\bf 73}, 107901 (2006).
\bibitem{az1}
O. Andreev and V.I. Zakharov, Phys.Rev.D {\bf 74}, 025023 (2006).
\bibitem{az4}
O. Andreev and V.I. Zakharov, Gluon Condensate, Wilson Loops and Gauge/String Duality, hep-ph/0703010.
\bibitem{az2}
O. Andreev and V.I. Zakharov, Phys.Lett.B {\bf 645}, 437 (2007).
\bibitem{az3}
O. Andreev and V.I. Zakharov, JHEP {\bf 04}, 100 (2007).
\bibitem{pisarski}
R.D. Pisarski, Fuzzy Bags and Wilson Lines, hep-ph/0612191; Phys.Rev.D {\bf 74}, 121703 (2006). 
\bibitem{karsch}
G. Boyd, J. Engels, F. Karsch, E. Laermann, C. Legeland, M. Lutgemeier, and B. Petersson, 
Phys.Rev.Lett. {\bf 75}, 4169 (1995); Nucl.Phys. B {\bf 469}, 419 (1996).
\bibitem{gavai}
R.V. Gavai, S. Gupta, and S. Mukherjee, Phys.Rev.D {\bf 71}, 0740013 (2005).
\bibitem{anomaly}
H. Leutwyler, in Proceedings of the Conference "QCD - 20 years later", World Scientific, p. 693.
\bibitem{mit}
A. Chodos, R.L. Jaffe, K. Johnson, C.B. Thorn, and V.F. Weisskopf, Phys.Rev.D {\bf 9}, 3471 (1974).
\bibitem{teper}
B. Bringoltz and M. Teper, Phys.Lett.B {\bf 628}, 113 (2005). 
\bibitem{teper2}
B. Lucini, M. Teper, and U. Wenger, JHEP {\bf 0401}, 061 (2004). 

\end{thebibliography}
\end{document}